# Time Aware Least Recent Used (TLRU) Cache Management Policy in ICN

Muhammad Bilal*, Shin-Gak Kang**

*Dept. of Engineering, University of Science and Technology (ETRI-Campus), Daejeon, Rep. of Korea
** Electronics and Telecommunications Research Institute (ETRI), Daejeon, Rep. of Korea
mbilal@etri.re.kr, sgkang@etri.re.kr

*Abstract*— The information centric networks (ICN) can be viewed as a network of caches. Conversely, ICN type of cache networks has distinctive features e.g, contents popularity, usability time of content and other factors inflicts some diverse requirements for cache eviction policies. In this paper we defined four important characteristics of a suitable eviction policy for ICN. We analysed well known eviction policies in view of defined characteristics. Based upon analysis we propose a new eviction scheme which is well suitable for ICN type of cache networks.

## I. INTRODUCTION

First time ICN gained the attention of research community when Van Jacobson identifies the basic paradigm shift in the internet services [1]. Internet trends are shifting away from browsing information to online consuming and sharing all types of content, including user generated contents. WHAT is being communicated is more important than WHO is communicating. ICN is a paradigm shift from host-centric to content-centric services and Source-driven to Receiver-driven approach. In ICN paradigm the network is then in charge of doing the mapping between the requested content and where it can be found. To do so Named Data Objects NDOs are used for identifying objects independent of its location or container. In network storage for caching NDOs is an integral part of the ICN service; hence ICN enabled network entities craft a network of caches. Nowadays all nodes in network potentially have memory to provide caching. Requests for NDOs can be satisfied by any node holding a copy in its cache. Hence, NDOs are located in the cache of authenticated and unauthenticated hosts/nodes scattered in the network.

The majority of research work is carried out to define proper architecture and system protocols (Naming, routing, congestion control etc) [1-8]. The performance analysis of ICN with respect to dynamics of in network caches has the primal importance. However, in literature very less work has been carried out to model and analyse cache dynamics of ICNs. There is a need to define analytical procedures and tools for the proper evaluation of ICN performance. In [9] authors introduced a network calculus for cache network however, the calculus toolkit is limited to LRU (Least Recently Used) replacement policy. In [10-13] authors introduced the analysis of cache network with TTL (Time To Live) based replacement policy. In [14-16] the performance analysis of LRU for cache networks discussed in detail. Thus, [17] provide an analytical toolkit for steady state performance analysis of in network caches. In this paper we propose a hybrid Time based LRU replacement scheme. The contents in ICN are different compared to web cache contents in term of availability and usability. We introduce a new terminology TTU (Time to Use), TTU is a time stamp of a content which stipulate the usability time for content based upon the locality of content and content publisher announcement. Owing to locality based time stamp, TTU provide more control to the local administrator to regulate in network storage.

## II. ANALYTICAL MODEL AND ASSUMPTIONS

Let us consider a network with *m* number of cache nodes and *k* number of contents. The content owner publishes content in network while a node generating a request for particular content is called the content consumer. A node can be a content publisher as well as a content consumer at the same time. While delivering content to a consumer, the intermediate cache nodes can store content for limited time duration. Any further request for that particular content can be fulfilled by intermediate cache nodes without reaching to publisher. A simple scenario is shown in Figure 1, consumer $n_a$ send a request for content $c_i$ at time $t$ (Solid thin line represent delivery route) while consumer $n_b$ and $n_c$ request same content at time $t+1$ (Dashed line represent delivery route). As a result consumer $n_b$ and $n_c$ get same content $c_i$ without visiting publisher node $n_p$. This practice greatly improves the content delivery time and reduces the network traffic.

**Notions**
- $TTU^i$ = Time stamp by the ith content publisher
- $TTU^i_j$ = Local Time stamp by jth cache node
- $N = \{n_1, n_2 \dots n_m\}$ = Set of cache node
- $|n_j|$ = Size of cache of $v_j$
- $C = \{c_1, c_2 \dots c_k\}$ Set of k content types in network
- $\lambda_{ij}$ = Exogenouse request rate of content $c_i$ at node $v_j$
- $\alpha_{ij}$ = Endogenouse request of content $c_i$ at node $v_j$
- $S = \{s[1], s[2] \dots s[m]\} = $ system state, where $s[j]$ represents the contents of $n_j$

- $\tau_{ij} = 1/E(r_{ij}) = $ The average time per request for $c_i \notin S[j]$ or in case of CCN $c_i \in PIT[j]$
- $\hat{\tau}_{ij} = 1/E(r_{ij}) = $ The average time per request for $c_i \in S[j]$ or for CCN $c_i \notin PIT[j]$

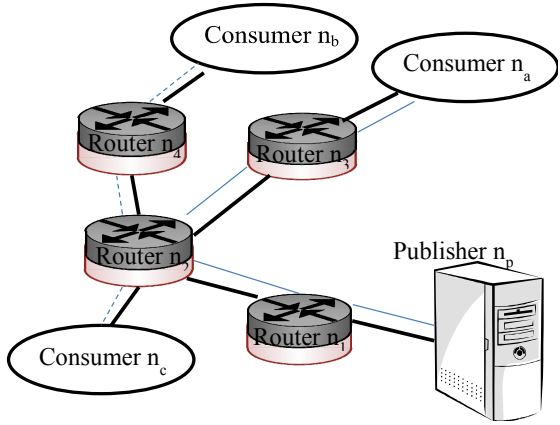

**Figure 1.** A simple cache network scenario

### A. In Network Content Copies

The storage for in network contents can be permanent or temporary. A permanent source node or publisher node $n_j$ keeps a permanent copy of content $(c_i)$. It has authority to modify the content, to publish the content and also it is responsible to assign/refresh/modify $TTU^i$ (A default $TTU^i = NULL$ which means don't store in cache and forward until destination node receives the content). A temporary source node or intermediate cache node $n_j$ keeps a temporary copy of content $(c_i)$. It has authority to forward content $(c_i)$ if $TTU_j^i > 0$. In case of publisher node $n_j$ and content $(c_i)$, we have $TTU_j^i = TTU^i$ where content availability is unconditional and content's $TTU_j^i/TTU^i$ doesn't change with passage of time. While in case of intermediate cache node $n_j$ and content $(c_i)$, we have $0 < TTU_j^i \leq TTU^i$ where content $TTU_j^i/TTU^i$ changes with passage of time i-e, content life span is $[0, TTU_j^i]$. Figure 2 shows the relationship between $TTU_j^i/TTU^i$ values for different level of cache nodes.

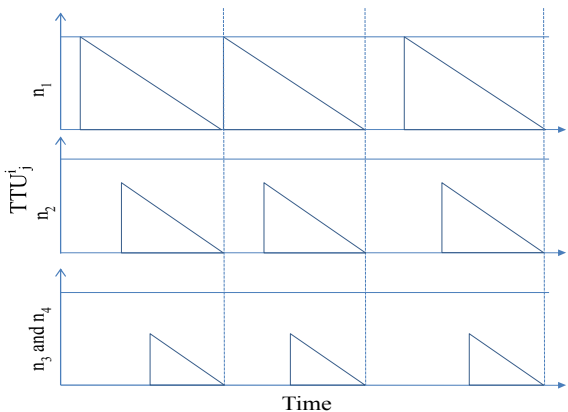

**Figure 2.** A simple cache network scenario

### B. Content Request Management

We define two types of content requests 1) Exogenous requests are the content requests generated by the consumer. A consumer can be end user or any other network entity. 2) Endogenous requests are those requests which are the miss rate from the lower cache nodes.

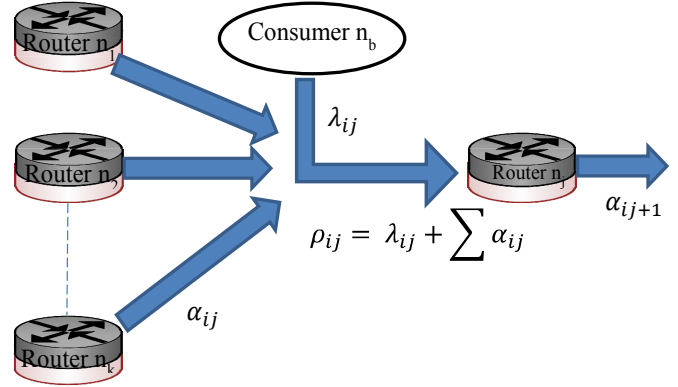

**Figure 3.** Exogenous and Endogenous content requests

As shown in Figure 3, $\lambda_{ij}$ is the exogenous request for content $i$ at cache node $n_j$, $\alpha_{ij}$ is the endogenous request for content $i$ at cache node $n_j$ and total number of requests can be found as below;

$$\rho_{ij} = \lambda_{ij} + \sum \alpha_{ij} \quad (1)$$

### C. Download Delay

For the simplicity, we consider zero download delay (ZDD) for requested content to be downloaded in cache node. The motive for this assumption is the observation that rate of arrival requests is smaller than the speed of downloading content.

### III. EVICTION POLICIES

In a long run cache get filled after certain amount of time. Upon arrival of new content when cache is already reached at full capacity an eviction policy replaces stored content with newly arrival content. A good eviction policy must have following characteristics.

- For the stability of the system an eviction policy must be ergodic (See the Definition-2)
- To reduce miss rate or increase hit rate, eviction policy should preserve the storage of most popular items available in cache.
- To keep more up to date items an eviction policy must be aware of expiration time of content availability.
- The computational complexity of eviction policy must be as simple as possible.

*Definition 1:* A cache state $s_k[j]$ is recurrent if the probability to return back to state $s_k[j]$ in finite time $(T_k)$ is 1, i-e $P_k[T_k] = P_k[T_k < \infty | s_0[j] = s_k[j]] = 1$.

*Definition 2:* A set of all recurrent cache states is called an ergodic set of cache.

*Definition 3:* An eviction policy is said to be non-protective if all contents in $s_k[j]$ at any given time $T_k$ has eviction probability $P_{ev}[s_k[j]] > 0 \ \forall \ c_i \in s_k[j]$.

According to [17] if an eviction policy is non-protective then it implies the ergodicity of eviction policy. Based upon the required characteristics of eviction policy we now discuss few well known eviction policies.

### A. First in First Out (FIFO)

FIFO is one the simplest and easy to implement eviction policy but for a large cache and huge content population size it is very hard to prove ergodicity of FIFO. FIFO is a protective eviction policy; hence it must not be possible to change the eviction order of cache state $s_k[j]$ at any given time $T_k$ except by visiting other cache states. For example in Figure 4 to change the evictions order $\{(a,b)(d,c)\} \rightarrow \{(a,b)(c,d)\}$ it requires 4 steps and possible only by changing the contents of cache, it gets more complex with increase of cache size. Moreover, FIFO can't conserve the popularity of contents and it is also unaware of expiration time of content availability.

### B. Least Recent Used and Least Frequent Used (LRU and LFU)

LRU and LFU are non-protective eviction policy; hence it is possible to change the eviction order of cache state $s_k[j]$ at any given time $(T_k)$ without changing the contents of cache, as shown in Figure 4 to move from one state to another it is possible without changing the contents of cache. LRU is also one the simplest and easy to implement eviction policy, but LRU can't well conserve the popularity of contents and it is also unaware of expiration time of content availability. On the other hand, LFU can conserve the popularity of contents but it is one the most complex and difficult to implement eviction policy, because LFU need to maintain the frequency of requests arrival for all contents. Moreover, LFU is also unaware of expiration time of content availability.

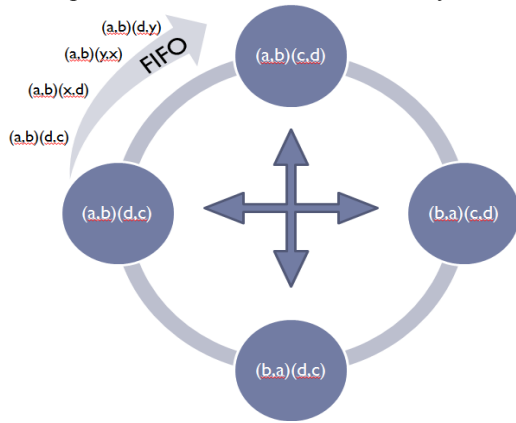

**Figure 4.** State Transition of LRU/LFU, arrow shows one state transition of FIFO for same contents

### C. Time Aware Least Recent Used (TLRU) Cache Management Policy

From above discussion it is clear that a more suitable eviction policy for ICN type of cache network is required. In this section we propose a new eviction scheme (TLRU) for ICN type of cache network. TLRU is an extension of simple LRU which is a non-protective (ergodic), content popularity and content life span aware eviction scheme. A brief stepwise explanation is given below.

**Step 1:** Calculate $TTU_j^i$ value for arriving content based upon composite function $f \uparrow g$ (Line 3). This step is optional and we argue that function should be defined by the local network administrator according to local policies and requirements. For our calculations we define function $f$ and $g$ intuitively, such that function $f$ maximizes the worth, if content size is large and minimizes if content size is small while function $g$ maximizes the worth, if content request is frequent and minimizes if content request rate is less frequent.

**Step 2:** Proceed to save arriving content in cache if the average request time $\tau_{ij}$ is smaller then $TTU_j^i$ calculated in step-1. The reasons for this step is that if the average request time (in CCN average request time can be calculated by using information stored in PIT-Pending Interest Table) $\tau_{ij} > TTU_j^i$ then there is a high probability that $TTU_j^i$ will expire before arrival of next request which means storing this content has no use. This step also endorses that relatively more popular contents should be stored.

**Step 3:** Store the content if there is an empty space in cache otherwise apply LRU on $Ev[j] \subseteq s_k[j]$. Subset $Ev[j]$ is a contraction of $s[j]$, calculated based upon the remaining $TTU$ value and average request time. Contraction endorses that relatively less popular contents should be evicted.

**Algorithm-(TLRU)**
1. $f = \frac{|c_i|}{|n_j|} * TTU^i$
2. $g = \frac{\rho_{ij}}{\sum_{k \neq i} \rho_{ij}} * TTU^i$
3. $TTU_j^i = f(TTU^i) \uparrow g(TTU^i)$
4. if $TTU_j^i > \tau_{ij}$
5.   if $s[j] \geq |n_j|$
6.     Do $\forall \ c_i \in s[j]$
7.       if $TTU_j^i < \hat{\tau}_{ij} \rightarrow c_i \in Ev[j]$
8.       $LRU(Ev[j]) \rightarrow Evict$
9.       $s[j] \rightarrow c_i \cup s[j]$
10.    else
11.      $LRU(s[j]) \rightarrow Evict$
12.      $s[j] \rightarrow c_i \cup s[j]$
13.  else $s[j] \rightarrow c_i \cup s[j]$
14. else $Reject$

## IV. ANALYSIS AND RESULTS

### A. Ergodicity

TLRU performs LRU eviction procedure on a contracted set $[j] \subseteq s_k[j]$. The contraction is a kind of filtering the less popular contents. We can consider it a time based random selection which implies that TLRU is non-protective; hence according to theorem-4 in [17] TLRU is ergodic.

## B. Queuing Delay

However, we considered zero download delay (ZDD) for requested content to be downloaded in cache node. The motive for this assumption is the observation that rate of arrival requests is smaller than the speed of downloading content. On the other hand, still a queuing delay exists in term of end user experience. If we consider each cache node as an M/M/1 queuing system, then mean waiting time of content is given by;

$$Mean\ Waiting\ time = \frac{\sigma}{T}\left[\frac{1/T}{1 - \sigma/T}\right] \quad (2)$$

Where $\sigma$ is mean contents arrival rate, then average network delay for $L$ number of intermediate hops can be calculated as;

$$Delay_{avg} = \sum_{i=1}^{L} \frac{\sigma/T}{T - T\sigma/T \cdot \prod_{j=1}^{i}(1 - h_p)} \quad (3)$$

## C. Eviction Time Approximation

For our analysis we defined a zipf-like popularity of content distribution as given below;

$$D(n) = \frac{1}{n^\alpha} \Rightarrow d(1) \geq d(2) \geq \cdots d(m) \quad (4)$$

Typical value of $\alpha$ is (0.6, 1.2).

Let's assume that $T_i$ is the cache eviction time for content $c_i$ and $T$ is the time required to populate cache to its maximum capacity, i-e, $|s[j]| = |n_j|$. Then by using Che approximation [18] we have $T_i = T$, which means the approximate eviction time for content $c_i$ is equal to time required to populate cache to its maximum capacity.

From the above analysis we conclude that a cache hit is possible if $T_i > \hat{\tau}_{ij}$ implies $T > \hat{\tau}_{ij}$. Now assuming Poisson arrival and assuming ZDD the probability of hit for TLRU is given by;

$$h_p = P(T > \hat{\tau}_{ij}) = E(1 - e^{-\rho(i,j)T}) \quad (5)$$

And successful probability of hit for LRU is given by;

$$\hat{h}_p = P(T > \hat{\tau}_{ij}).P(TTU_j^i > \tau_{ij}) \quad (6)$$

Consequently we get,

$$|s[j]| = |n_j| = Const = \sum_{k=1, k \neq i}^{m} (1 - e^{-\rho(i,j)T}) \quad (7)$$

In the above equation $T$ is unknown. To find out value of $T$ Equation 7 can be solved by using "newton method". Let's define a function $F(T)$;

$$F(T) = Const - \sum_{k=1, k \neq i}^{m} (1 - e^{-\rho(i,j)T}) \quad (8)$$

To start newton iteration we have to guess initial value of $T$, let the following intuitive equation is our initial guess;

$$T = \frac{Const}{\sum_{k=1, k \neq i}^{m}(\rho_{ij})} \quad (9)$$

After initial guess now we perform newton iteration (Given in Equation 10) until the value of $T$ converges to some constant value as shown in Figure 5 we get a constant value after 3rd iteration;

$$T_{new} = T_{old} - \frac{F(T_{old})}{\acute{F}(T_{old})} \quad (10)$$

Where $\acute{F}(T)$ is the derivative if function $F(T)$ defined in Equation 8.

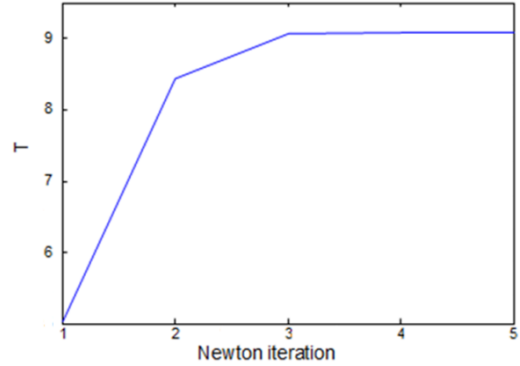

**Figure 5.** Value of T converges to constant value after 3rd newton iteration

## D. Performance Comparison Results

Let's consider $10^4$ numbers of contents in a network. The popularity of contents is determined by using Zipf-distribution given in Equation 4. Moreover, let's assume *TTU* value is normally distributed. After finding the value of $T$ using newton method (as discussed in Section-IV-C), we can get numerical results of probability of hit for TLRU and LRU. The results are taken for different cache size, shown in Figure 6 to 8. It is clear that for the small cache size, performance gap of LRU and TLRU is very small. However, the performance gap increases with the increase of cache size.

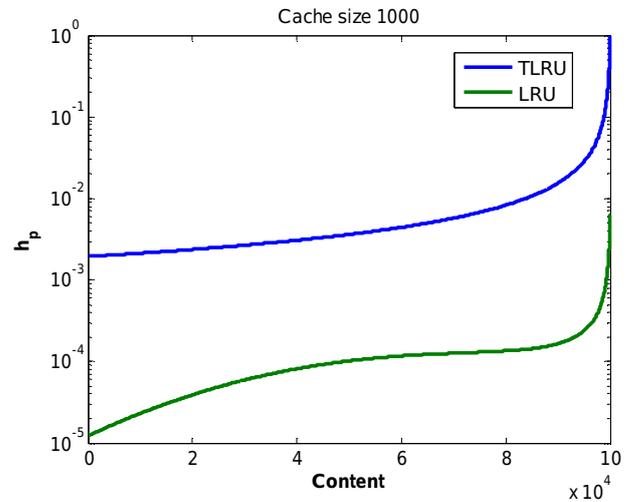

**Figure 6.** TLRU and LRU probability of hit for cache size of 1000

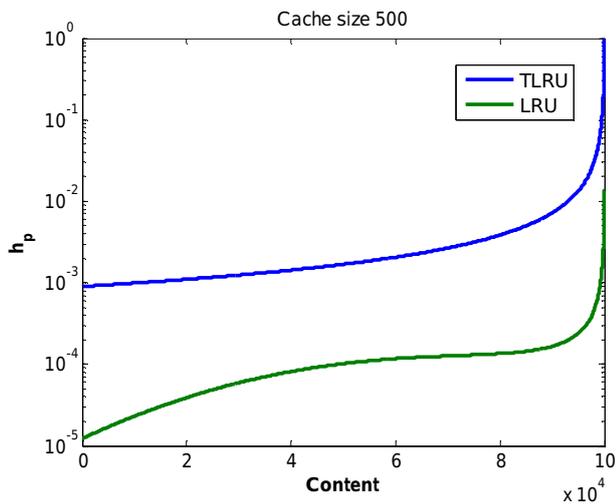

**Figure 7.** TLRU and LRU probability of hit for cache size of 500

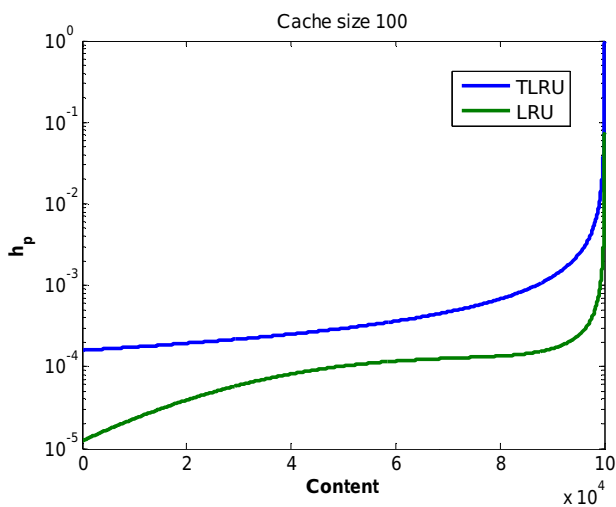

**Figure 8.** TLRU and LRU probability of hit for cache size of 100

## V. CONCLUSIONS

In this work we defined four important characteristics of a suitable eviction policy for ICN type of cache networks. We analyzed well known eviction policies (FIFO, LRU and LFU) to check the suitablity for cache networks and found that these eviction policies has variouse draw backs to handle content replacement in ICN type of cache networks. To overcome these drawbacks we proposed a new scheme TLRU which is an extended version of LRU. TLRU take account of 'content usable time' along with lagacy LRU scheme such that it conserve to store high popular contents and as a result is increases probability of hit.

## ACKNOWLEDGMENT

This research was supported by the ICT Standardization program of MSIP(The Ministry of Science, ICT & Future Planning).

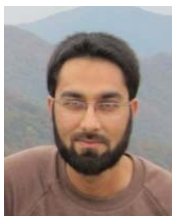

**Muhammad Bilal** has received his BS degree in Computer Systems Engineering from University of Engineering and Technology (UET) Peshawar Pakistan and MS in Computer Engineering from Chosun University Korea. Currently he is Ph.D student at University of Science and Technology (UST) Korea in Electronics and Telecommunication Research Institute (ETRI) campus. His research interests include Design and Analysis of Network Protocols, Network Architecture and Future Internet.

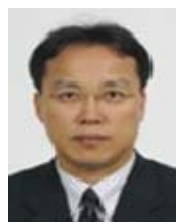

**Shin-Gak Kang** received the B.S. and M.S. degree in electronics engineering from Chungnam National University, Korea, in 1984 and 1987, respectively and the Ph.D. degree in information communication engineering from Chungnam National University, Korea in 1998. He is working for ETRI since 1984. Currently, he is a Director of media application standard research section. From 2008 he is a professor at the Department of Broadband Network Technology, University of Science and technology, Korea. He is actively participating in various international standard bodies as a Vice-chairman of ITU-T SG11, Convenor of JTC 1/SC 6/WG 7, etc. His research interests include multimedia communications, contents networking, and Future Network.